\documentclass[aps,twocolumn,prl,showpacs,showkeys,nofootinbib,superscriptaddress]{revtex4-1}
\usepackage{amsmath, amssymb, amsfonts}
\usepackage[margin=1in]{geometry}
\usepackage{graphicx}
\usepackage{color}
\usepackage{xcolor}
\usepackage{lineno}
\usepackage{bm}
\usepackage{ulem}

\newcommand{\bea}{\begin{eqnarray}}
\newcommand{\eea}{\end{eqnarray}}
\newcommand{\beq}{\begin{equation}}
\newcommand{\eeq}{\end{equation}}
\newcommand{\bqa}{\begin{eqnarray}}
\newcommand{\eqa}{\end{eqnarray}}

\graphicspath{{Figures/}}

\begin{document}
\title{Evolution of charm-meson ratios \\in an expanding hadron gas}
\author{Eric Braaten}
\email{braaten.1@osu.edu}
\affiliation{Department of Physics,
	         The Ohio State University, Columbus, OH\ 43210, USA}
	
\author{Roberto Bruschini}
\email{roberto.bruschini@ific.uv.es}
\affiliation{
Instituto de F\'isica Corpuscular, Universidad de Valencia y Consejo Superior de Investigationes Cient\'ificas, 
46980 Paterna, Spain}
	
\author{Li-Ping He}
\email{heliping@hiskp.uni-bonn.de}
\affiliation{Helmholtz-Institut f\"ur Strahlen- und Kernphysik and Bethe Center for Theoretical Physics, Universit\"at Bonn, D-53115 Bonn, Germany}
	
\author{Kevin Ingles}
\email{ingles.27@buckeyemail.osu.edu}
\affiliation{Department of Physics,
	         The Ohio State University, Columbus, OH\ 43210, USA}
	
\author{Jun Jiang}
\email{jiangjun87@sdu.edu.cn}
\affiliation{School of Physics, Shandong University, Jinan, Shandong 250100, China}
\date{\today}

\begin{abstract}
We study the time evolution of the numbers of charm mesons 
after the kinetic freeze-out of the hadron gas produced by a central heavy-ion collision. 
The $\pi D^\ast \to \pi D^\ast$ reaction rates have $t$-channel singularities 
that give contributions inversely proportional to the thermal width of the $D$.
The ratio of the $D^0$ and $D^+$ production rates
can differ significantly from those predicted using the measured $D^\ast$ branching fractions.
\end{abstract}


\maketitle
	
\textbf{Introduction.}
A remarkable aspect of the pseudoscalar and 
vector charm mesons $D$ and $D^\ast$ is that the $D^\ast$-$D$ mass splittings 
are all very close to the pion mass $m_\pi$.
A previously unrecognized consequence is
that there are charm-meson reactions with $t$-channel singularities.
A $t$-channel singularity is a divergence in the rate for a reaction in which an unstable
particle decays and one of its decay products is scattered.
The singularity arises if the exchanged particle can be on shell.
The existence of $t$-channel singularities was first pointed out by Peierls in 1961
in the context of $\pi N^\ast$ scattering \cite{Peierls:1961zz}.
An example in the Standard Model of particle physics is the reaction  $\mu^+ \mu^- \to W^+ e^- \bar \nu_e$, 
which can proceed through exchange of $\nu_\mu$.
Melnikov and Serbo solved the divergence problem by taking into account
 the finite transverse sizes of the colliding $\mu^+$ and $\mu^-$ beams \cite{Melnikov:1996na}.
A general discussion of  $t$-channel singularities was presented by 
Grzadkowski, Iglicki, and Mr\'owczy\'nski \cite{Grzadkowski:2021kgi}.
They pointed out that in a thermal medium, 
a $t$-channel singularity is regularized by the thermal width of the exchanged particle.

The simplest charm-meson reactions with a $t$-channel singularity are $\pi D^\ast \to \pi D^\ast$,
which can proceed through the decay $D^\ast \to D \pi$ followed by the 
inverse decay $\pi D\to D^\ast$.
The $t$-channel singularity arises because the exchanged $D$ can be on shell.
The cross section diverges
when the square of the center-of-mass energy $s$ is in a narrow interval near the threshold.
In the case of elastic scattering, 
the $t$-channel singularity region is
\beq
2 M_*^2 - M^2 + 2 m^2 < s < (M_*^2 - m^2)^2/M^2,
\label{t-sing:piD}
\eeq
where $M_*$, $M$, and $m$ are the masses of $D^\ast$, $D$, and $\pi$.
The interval in $\sqrt{s}$ is largest for the reaction $\pi^0 D^{\ast 0} \to \pi^0 D^{\ast 0}$,
extending from 6.1~MeV to 8.1~MeV above the threshold $M_*+m=2141.8$~MeV. 

An obvious question is whether $t$-channel singularities in charm-meson reactions 
have any observable consequences.
One possibility is that $t$-channel singularities could modify the ratios of charm mesons 
produced in a high-energy collision.
We denote the numbers of $D^0$, $D^+$, $D^{\ast 0}$, and $D^{\ast +}$
by $N_0$, $N_+$, $N_{\ast 0}$, and $N_{\ast +}$.
The observed numbers of $D^0$  and $D^+$ can be predicted in terms of the 
numbers $(N_a)_0$ and $(N_{\ast a})_0$ before $D^\ast$ decays 
and the measured branching fraction,
$B_{+0} = (67.6 \pm 0.5)\%$ for $D^{\ast +} \to D^0 \pi^+$:
\begin{subequations}
\begin{align}
N_0 &= 
\left( N_0 \right)_0  + \left( N_{\ast 0} \right)_0  + B_{+0}  \left( N_{\ast +} \right)_0 ,
\label{ND0}
\\
N_+ &= 
\left( N_+ \right)_0 + 0  +  (1 - B_{+0}) \left( N_{\ast +} \right)_0 ,
\label{ND+}
\end{align}
\label{ND0,p}%
\end{subequations}
where  the last two terms come from $D^{\ast 0}$ and $D^{\ast +}$ decays, respectively.
These simple relations have been assumed in all previous analyses of charm-meson production.
We will show that Eqs.~\eqref{ND0,p} can be modified by $t$-channel singularities.

One situation in which the effects of $t$-channel singularities
may be observable is in the production of charm mesons  from ultrarelativistic heavy-ion collisions.
A sufficiently central heavy-ion collision is believed to produce a region of quark-gluon plasma (QGP)
that makes a transition to a hadron resonance gas (HRG).
In the HRG, $t$-channel singularities are regularized by the thermal widths of the hadrons.
The most divergent term in a reaction rate is replaced by a term inversely proportional to the thermal width
of the exchanged hadron.

In this paper, we study the effects of $t$-channel singularities in charm-meson reactions 
in the expansion of the hadron gas after kinetic freeze-out.
The restriction to after kinetic freeze-out offers many simplifications. 
The only  hadrons in the HRG that remain are the most stable ones with lifetimes $c\tau \gg 1$~fm.
The temperature is low enough that the important interactions can be described by an appropriate effective field theory.
The only relevant charm mesons are $D$ and  $D^\ast$.
The decays of $D^\ast$ occur at much later times ($c\tau > 2000$~fm).
The most abundant hadrons by far are pions,
so the dominant contribution to the thermal width of  a charm meson 
comes from the forward scattering of pions
and is proportional to the decreasing pion number density.

\textbf{Charm mesons.}
We denote the masses of $D^+$,
$D^0$, $D^{\ast+}$, and $D^{\ast0}$ by $M_+$, $M_0$, $M_{\ast+}$, and $M_{\ast0}$.
We denote the masses of $\pi^+$ and $\pi^0$ by $m_{\pi +}$ and $m_{\pi 0}$ (or collectively by $m_\pi$).
The decay width of $D^{\ast+}$ is measured: $\Gamma_{\ast+} = 83.4 \pm 1.8$~keV.
The decay width of $D^{\ast0}$ can be predicted using Lorentz invariance, 
chiral symmetry, isospin symmetry, and the measured $D^\ast$ branching fractions:
$\Gamma_{\ast0} = 55.4 \pm 1.5$~keV.
The radiative decay rates of $D^{\ast 0}$ and $D^{\ast +}$ are 
$\Gamma_{*0,\gamma} = 19.6 \pm 0.7$~keV and $\Gamma_{*+,\gamma} = 1.3 \pm 0.3$~keV. 

Pions that are not ultrarelativistic can be described by chiral effective field theory ($\chi$EFT).
The self-interactions of pions in $\chi$EFT at leading order (LO) 
are determined by the pion decay constant $f_\pi = 131.7$~MeV. 
Interactions between charm mesons and pions can be described using heavy hadron $\chi$EFT (HH$\chi$EFT)
\cite{Burdman:1992gh,Wise:1992hn,Cheng:1992xi}.
At LO, the interactions are determined by $f_\pi$
and a dimensionless coupling constant $g_\pi = 0.5204 \pm 0.0059$.

In heavy-ion collisions,
the QGP can have a substantial effect on the production of charm hadrons \cite{Dong:2019byy}.
A simple model for particle production in heavy-ion collisions is the Statistical Hadronization Model 
(SHM)~\cite{Andronic:2005yp}.
According to SHM, light hadrons are produced during hadronization, 
with the QGP and the HRG in thermal and chemical equilibrium. 
The application of SHM to charm hadron production is sometimes called SHMc \cite{Andronic:2021erx}.
Charm quarks are primarily created in the hard collisions of the heavy ions. 
They are assumed to be in thermal equilibrium with the QGP,
but they remain out of chemical equilibrium,
since the charm-quark mass is much larger than the temperature $T$. 
Conservation of charm-quark number determines the charm-quark fugacity as a function of $T$.
According to SHMc, charm hadrons are produced during hadronization,
with the charm quarks and charm hadrons
in thermal equilibrium with the same fugacity $g_c$.
As the HRG expands and cools, the total number of charm mesons remains essentially  constant.
After chemical freeze-out, the abundances of more stable charm hadrons
are increased by the decays of higher charm-hadron resonances.

For Pb-Pb collisions at nucleon-nucleon center-of-mass energy $\sqrt{s_{NN}}=5.02$~TeV, 
the charm-quark fugacity at hadronization has been determined to be $g_c = 29.6 \pm 5.2$~\cite{Andronic:2021erx}.
The predictions of SHMc for the multiplicities of charm hadrons are given in Ref.~\cite{Andronic:2021erx}.
The predicted multiplicities $dN/dy$ at midrapidity ($|y| < \tfrac12$) from collisions in the centrality range 0-10\% 
are 6.42 for $D^0$, 2.84 for $D^+$, and 2.52 for $D^{\ast +}$, with errors that come predominantly from $g_c$.
We estimate the multiplicity of $D^{\ast 0}$ by assuming 
$D^{\ast 0}$ and  $D^{\ast +}$ are in thermal equilibrium at kinetic freezeout 
with temperature $T_F=115$~MeV: 
$N_{\ast 0}/N_{\ast +} = (M_{\ast 0}/M_{\ast +})^2 K_2(M_{\ast 0}/T_F)/ K_2(M_{\ast +}/T_F)$.
The estimated multiplicity for $D^{\ast 0}$ is 2.59. 
We obtain  plausible  multiplicities before $D^*$ decays by using Eqs.~\eqref{ND0,p}
with $B_{+0} = 67.7$: 
\begin{subequations}
\begin{align}
(dN_0/dy)_0 &= 2.12, &(dN_+/dy)_0 & = 2.03, 
\\
(dN_{\ast 0}/dy)_0 &= 2.59, & (dN_{\ast +}/dy)_0 &= 2.52.
\end{align}
\label{dND0,dNDp}%
\end{subequations}
Given these multiplicities,
the ratio $N_0/N_+$ of the $D^0$ and $D^+$ multiplicities is predicted by Eqs.~\eqref{ND0,p} to increase 
from 1.044 before $D^*$ decays  to $2.256 \pm 0.014$ 
at the detector, where the error bar is from $B_{+0}$ only.

\textbf{Expanding hadron gas.}
A simple model for the system produced by a sufficiently central heavy-ion collision
at a proper time $\tau$ after the collision is
a homogeneous system with volume $V(\tau)$ in thermal equilibrium at a temperature $T(\tau)$.
Parametrizations of $V(\tau)$ and $T(\tau)$ 
based on the boost-invariant longitudinal expansion proposed by Bjorken~\cite{Bjorken:1982qr}
and an accelerated transverse expansion were presented in Ref.~\cite{Hong:2018mpk}.
After kinetic freeze-out at proper time $\tau_F$, 
the system continues to expand, but its temperature remains fixed at $T_F$.
A simple model for the volume $V(\tau)$ for $\tau>\tau_F$ is continued longitudinal expansion at the speed 
of light and transverse expansion at the same speed $v_F$ as at kinetic freeze-out:
\begin{align}
V(\tau) = \pi \left[ R_F + v_F(\tau - \tau_F) \right]^2 c\tau,
\label{V-tau>}
\end{align}
where $R_F$ is the transverse  radius at kinetic freezeout.
The parameters can be determined by fitting the output of a simplified hydrodynamic model \cite{Hong:2018mpk}.
The parameters for Pb-Pb collisions at 5.02~TeV are $\tau_F = 21.5$~fm/$c$, $R_F = 24.0$~fm,
and $v_F = 1.00~c$ \cite{Abreu:2020ony}.
The kinetic freeze-out temperature is $T_F=115$~MeV.

The number density for pions at kinetic freeze-out $\mathfrak n_\pi(\tau_F)$ can be
estimated by assuming they are an ideal gas of relativistic bosons in thermal and chemical
equilibrium at temperature $T_F$.
At later times, $\mathfrak n_\pi(\tau)$ is decreased by the expansion of the system:
$\mathfrak{n}_\pi(\tau) = [V(\tau_F)/V(\tau)]\mathfrak{n}_\pi(\tau_F)$.
The multiplicities of $\pi^+$ and $\pi^-$ produced by Pb-Pb collisions at the LHC 
have been measured by the ALICE collaboration at $\sqrt{s_{NN}}=5.02$~TeV \cite{ALICE:2019hno}.
The common pion multiplicity for $\pi^+$, $\pi^-$, and $\pi^0$ in the centrality range 0-10\% 
is $dN_\pi/dy=769 \pm 34$.
We can use the pion multiplicity and the predicted charm-meson multiplicities in Eqs.~\eqref{dND0,dNDp}
to estimate the charm-meson number densities at times $\tau$  before $D^*$ decays:
\beq
\mathfrak{n}_{D^{(\ast)}}(\tau) = 
[(dN_{D^{(\ast)}}/dy)/(dN_\pi/dy)]_0\, \mathfrak{n}_{\pi}(\tau).
\label{nD,D*-F}
\eeq

\textbf{Thermal masses and widths.}
At kinetic freeze-out, the pion momentum distribution $\mathfrak{f}_\pi$ 
is a Bose-Einstein distribution:
$\mathfrak f^{(\mathrm{eq})}_\pi(\omega_q) = 1/(e^{\omega_q/T_F} -1)$,
where $\omega_q = \sqrt{m_\pi^2 + q^2}$ and $T_F=115$~MeV.
The pion number density is $\mathfrak{n}_\pi^\mathrm{(eq)} = \int \mathfrak f^{(\mathrm{eq})}_\pi(\omega_q)d^3q/(2\pi)^3$.
We assume the subsequent expansion of the pion gas is isothermal, so the pion momentum distribution for $\tau > \tau_F$  is 
$\mathfrak{f}_\pi(\omega_q)= 
(\mathfrak{n}_\pi(\tau)/\mathfrak{n}_\pi^\mathrm{(eq)})\, \mathfrak f^{(\mathrm{eq})}_\pi(\omega_q)$.
We assume $D^+$, $D^0$, $D^{\ast+}$, and $D^{\ast0}$ 
all have relativistic Boltzmann distributions with the same temperature $T_F$.

When a particle propagates through a medium, 
its properties are modified by the interactions with the medium. 
The thermal mass shift and thermal width of a particle
can be obtained by evaluating its in-medium self-energy on the mass shell and at zero 3-momentum. 
The pion mass shift and thermal width after kinetic freeze-out
can be calculated using $\chi$EFT at LO. The pion mass shift is \cite{Gasser:1986vb}
\begin{equation}
\delta m_\pi= 
(m_\pi/2 f_\pi^2)\, \mathfrak{n}_\pi \left\langle 1/\omega_q \right\rangle ,
\label{deltampi}
\end{equation}
where the angular brackets represents the thermal average over the pion momentum $\bm q$.
The pion thermal width is 0 at this order.
For pions with the equilibrium number density
$\mathfrak{n}_\pi^\mathrm{(eq)}$ at $T_F= 115$~MeV,
the mass shift is $\delta m_\pi=1.550$~MeV. 
The thermal mass shift and thermal width of a charm meson can be calculated using HH$\chi$EFT at LO.
The $D$ and $D^\ast$ mass shifts are insensitive to isospin splittings:
\begin{equation}
\delta M =
(3g_\pi^2/2f_\pi^2)\, \mathfrak{n}_\pi \, \Delta
\left \langle 1/\omega_q \right\rangle,
~
\delta M_\ast = - \delta M/3,
\label{deltaM-approx}
\end{equation}
where $\Delta =141.3$~MeV is the average of the four $M_* - M$ mass splittings.
For pions with $\mathfrak{n}_\pi^\mathrm{(eq)}$ and $T_F= 115$~MeV,
the $D$ thermal mass shift is $\delta M = 1.296$~MeV. 
The thermal widths of $D$ and $D^\ast$ are sensitive to isospin splittings only through the
rates $\Gamma_{\ast b,d}$ for the decays $D^{\ast b}\to D^d \pi$.
The decay rates in the vacuum are 
\begin{subequations}
\begin{align}
\Gamma_{{\ast +},+} &= \frac{g_\pi^2}{12\pi \, f_\pi^2} 
\left[ (M_{\ast +}- M_{+})^2 - m_{\pi 0}^2 \right]^{3/2},
\label{GammaD*+D+pi}
\\
\Gamma_{{\ast +},0} &= \frac{g_\pi^2}{6\pi \, f_\pi^2} 
\left[ (M_{\ast +}- M_{0})^2 - m_{\pi +}^2 \right]^{3/2},
\label{GammaD*pD0pi}
\\
\Gamma_{{\ast 0},0} &= 
\frac{g_\pi^2}{12\pi \, f_\pi^2} 
\left[ (M_{\ast 0}- M_{0})^2 - m_{\pi 0}^2 \right]^{3/2},
\label{GammaD*0D0pi}
\end{align}
\label{GammaD*Dpi}%
\end{subequations}
and $\Gamma_{\ast0,+} = 0$.
In the hadron gas, these rates are modified by
the thermal mass shifts in Eqs.~\eqref{deltampi} and \eqref{deltaM-approx}.
The thermal mass shifts decrease the available phase space, so
the decay $D^{\ast 0}\to D^+ \pi^-$  remains kinematically  forbidden
for $T=T_F$ and $\mathfrak{n}_\pi < \mathfrak{n}_\pi^\mathrm{(eq)}$.
The  interaction widths of $D^a$ and $D^{\ast a}$ in the hadron gas are
\begin{subequations}
\begin{align}
\Gamma_a &=
3 \, \mathfrak f_\pi(\Delta) \sum_c \Gamma_{\ast c,a},
\\
\Gamma_{\ast a} &= 
 [1+\mathfrak f_\pi(\Delta) ]\sum_c \Gamma_{\ast a,c} + \Gamma_{*a,\gamma} , 
\label{deltaGamma-approx}
\end{align}
\end{subequations}
where $\Gamma_{*a,\gamma}$ is the radiative decay rate of $D^{\ast a}$.
The thermal contribution has a factor $\mathfrak f_\pi(\Delta)$,
which is proportional to $\mathfrak n_\pi$.

\textbf{Reaction Rates.}
The reaction rates $\langle v\sigma_{\pi a,\ast b}\rangle $  in HH$\chi$EFT at LO 
for the inverse decay $\pi D^a  \to D^{\ast b}$, averaged over the 3 pion isospins, are
\begin{subequations}
\begin{align}
\left\langle v\sigma_{\pi + , {\ast +}} \right\rangle 
&= \big[ \mathfrak{f}_\pi(\Delta)/\mathfrak{n}_\pi\big]\, 
\Gamma_{{\ast +} , +},
\label{vsigmaD*+D+pi}
\\
\left\langle v\sigma_{\pi 0 , {\ast +}} \right\rangle 
&=\big[ \mathfrak{f}_\pi(\Delta)/\mathfrak{n}_\pi\big]\, 
\Gamma_{{\ast +} ,0} ,
\label{vsigmaD*pD0pi}
\\
\left\langle v\sigma_{\pi 0 , {\ast 0}} \right\rangle 
&= \big[ \mathfrak{f}_\pi(\Delta)/\mathfrak{n}_\pi \big]\, 
\Gamma_{{\ast 0} , 0} ,
\label{vsigmaD*0D0pi}
\end{align}
\label{vsigmaD*Dpi}%
\end{subequations}
and $\langle v\sigma_{\pi + , {\ast 0}} \rangle = 0$.
Note that the factor $\mathfrak{f}_\pi(\Delta)/\mathfrak{n}_\pi 
= \mathfrak{f}_\pi^\mathrm{(eq)}(\Delta)/\mathfrak{n}_\pi^\mathrm{(eq)}$ does not depend on 
the pion number density, so the dependence on $\mathfrak{n}_\pi$ enters 
only through the mass shifts in $\Gamma_{\ast a,b}$ given in Eqs.~\eqref{GammaD*Dpi}.
The rates for the reactions $\pi D^\ast \leftrightarrow \pi D$, which change the charm-meson spin,
can be simplified by taking the limit $M_* - M \to m_\pi$.
The reaction rates $\langle v\sigma_{\pi \ast a,\pi b}\rangle $ for $\pi D^{*a} \to \pi D^b$ 
and $\langle v\sigma_{\pi a,\pi \ast b}\rangle $ for $\pi D^a \to \pi D^{*b}$ at $T_F$,
averaged over initial and final pion isospins, are
\begin{subequations}
\begin{align}
\left\langle v\sigma_{\pi *0, \pi 0} \right\rangle = \left\langle v\sigma_{\pi *+, \pi +} \right\rangle 
&= 0.2446\, g_\pi^4\, m_\pi^2/f_\pi^4  ,
\label{vsigmapiD*0D0}
\\
\left\langle v\sigma_{\pi *0, \pi +} \right\rangle =\left\langle v\sigma_{\pi *+, \pi 0} \right\rangle 
&= 0.0056\, g_\pi^4\, m_\pi^2/f_\pi^4,
\label{vsigmapiD*0D+}
\\
\left\langle v\sigma_{\pi 0, \pi *0} \right\rangle = \left\langle v\sigma_{\pi +, \pi *+} \right\rangle 
&= 0.2181\,g_\pi^4\, m_\pi^2/f_\pi^4 , 
\label{vsigmapiD0D*0-old}
\\
\left\langle v\sigma_{\pi 0, \pi *+} \right\rangle 
=\left\langle v\sigma_{\pi +, \pi *0} \right\rangle 
&= 0.0049\, g_\pi^4\, m_\pi^2/f_\pi^4 .
\label{vsigmapiD0D*+-old}
\end{align}
\label{vsigmapiDD*}%
\end{subequations}
The numerical factors depend only on $m_\pi/T_F$.
We have used $m_\pi = (m_{\pi 0} + m_{\pi +}) / 2$ and $T_F=115$~MeV.

The reaction $\pi D^a  \to \pi D^{b}$ has a resonance from an intermediate $D^{\ast c}$ in the $s$ channel.
Its reaction rate $\langle v\sigma_{\pi a,\pi b}\rangle$ can be approximated by the sum of
the nonresonant contribution defined by the limit $M_* - M \to m_\pi$
and the resonant contribution proportional to $1/\Gamma_{\ast c}$.
The reaction $\pi D^{\ast a}  \to \pi D^{\ast b}$ has a $t$-channel singularity from an intermediate $D^{c}$.
Its  reaction rate $\langle v\sigma_{\pi \ast a,\pi \ast b}\rangle $ can be approximated by the sum of
the nonsingular contribution defined by the limit $M_* - M \to m_\pi$
and the most singular contribution proportional to $1/\Gamma_{c}$.
The  rates  at $T_F$ for the reactions that change the charm-meson flavor,
averaged over initial and final pion isospins,  are
\begin{widetext}
\begin{subequations}
\begin{align}
\left\langle v\sigma_{\pi 0 , \pi +} \right\rangle 
= \left\langle v\sigma_{\pi + , \pi 0} \right\rangle & = 
(1.0007 + 0.3336 \,g_\pi^4)\, \frac{m_\pi^2}{ f_\pi^4} 
+ \frac{\mathfrak{f}_\pi(\Delta)}{\mathfrak{n}_\pi} \,
\frac{\Gamma_{{\ast +} , 0}\, \Gamma_{{\ast +} , +} }{\Gamma_{*+}},
\label{vsigmapiD0+,+0}
\\
\left\langle v\sigma_{\pi *0, \pi *+} \right\rangle 
= \left\langle v\sigma_{\pi *+, \pi *0} \right\rangle & =
(1.0007 + 0.3086\, g_\pi^4) \frac{m_\pi^2}{ f_\pi^4}  
+ \frac{\mathfrak{f}_\pi(\Delta)}{\mathfrak{n}_\pi}\,
\frac{\Gamma_{\ast 0,0} \, \Gamma_{\ast +,0}}{\Gamma_0} .
\label{vsigmapiD*0D*+,+0}
\end{align}
\end{subequations}
\end{widetext}
The dependence on $\mathfrak{n}_\pi$ enters through the mass shifts in $\Gamma_{\ast a,b}$ 
and  the factors of $1/\Gamma_{*+}$ and $1/\Gamma_0$.
The resonance term in Eq.~\eqref{vsigmapiD0+,+0} is 
about three  orders of magnitude smaller than  the nonresonant term for $n_\pi < n_\pi^\mathrm{(eq)}$.
The $t$-channel singularity term in Eq.~\eqref{vsigmapiD*0D*+,+0} is larger than the nonsingular term 
when $n_\pi < 10^{-3}\,  n_\pi^\mathrm{(eq)}$. 

\textbf{Evolution Equations.}
The time-evolution equation for a charm meson in the expanding hadron gas
is most conveniently expressed as an equation for the ratio of 
 its number density to the pion number density $n_\pi$; this
removes the effect of the increasing volume of the hadron gas.
The evolution equations for $D^a$ and $D^{*a}$ are 
\begin{widetext}
\begin{subequations}
\begin{multline}
\mathfrak{n}_\pi \frac{d \ }{d\tau} \left( \frac{\mathfrak{n}_{D^a}}{\mathfrak{n}_\pi} \right) =
[1+\mathfrak f_\pi(\Delta) ] \sum_b \Gamma_{\ast b,a} \,\mathfrak{n}_{D^{\ast b}} 
+\Gamma_{\ast a,\gamma} \,\mathfrak{n}_{D^{\ast a}}
- 3\sum_b \left\langle v\sigma_{\pi a ,\ast b} \right\rangle \, \mathfrak{n}_{D^a}\, \mathfrak{n}_{\pi} 
\\
+ 3\sum_{b \neq a} \left\langle v\sigma_{\pi b, \pi a} \right\rangle \, 
\big( \mathfrak{n}_{D^b}
- \mathfrak{n}_{D^a} \big) \mathfrak{n}_\pi
+ 3\sum_b  \big( \left \langle v\sigma_{\pi \ast b, \pi a} \right\rangle \,\mathfrak{n}_{D^{\ast b}} 
- \left \langle v\sigma_{\pi a, \pi \ast b} \right\rangle \mathfrak{n}_{D^a} \big) \, \mathfrak{n}_{\pi} 
+\ldots,
\label{dnDa/dt}
\end{multline}
\begin{multline}
\mathfrak{n}_\pi \frac{d \ }{d \tau} \left( \frac{\mathfrak{n}_{D^{\ast a}}}{\mathfrak{n}_\pi} \right) =
3 \sum_b \left\langle v\sigma_{\pi b \to {\ast a}} \right\rangle \mathfrak{n}_{D^b}\, \mathfrak{n}_\pi
 - \Big( [1+\mathfrak f_\pi(\Delta) ] \sum_b \Gamma_{\ast a,b} + \Gamma_{\ast a,\gamma} \Big) \mathfrak{n}_{D^{\ast a}}
\\
+ 3 \sum_b  \big( \left \langle v\sigma_{\pi b, \pi \ast a} \right\rangle \,\mathfrak{n}_{D^b} 
- \left \langle v\sigma_{\pi \ast a, \pi b} \right\rangle \mathfrak{n}_{D^{*a}} \big) \mathfrak{n}_\pi
+ 3 \sum_{b \neq a} \left\langle v\sigma_{\pi *b, \pi \ast a} \right\rangle \, \big( \mathfrak{n}_{D^{*b}}
-  \mathfrak{n}_{D^{\ast a}} \big) \mathfrak{n}_\pi
+\ldots.
\label{dnD*a/dt}
\end{multline}
\label{dnD,D*a/dt}
\end{subequations}
\end{widetext}
The ratio of the total charm-meson density
 $\mathfrak{n}_{D^0} + \mathfrak{n}_{D^+} + \mathfrak{n}_{D^{*0}} + \mathfrak{n}_{D^{*+}}$
and $\mathfrak n_\pi$ remains constant in accordance with charm-quark conservation.

\begin{figure}[t]
\includegraphics[width=0.49\textwidth]{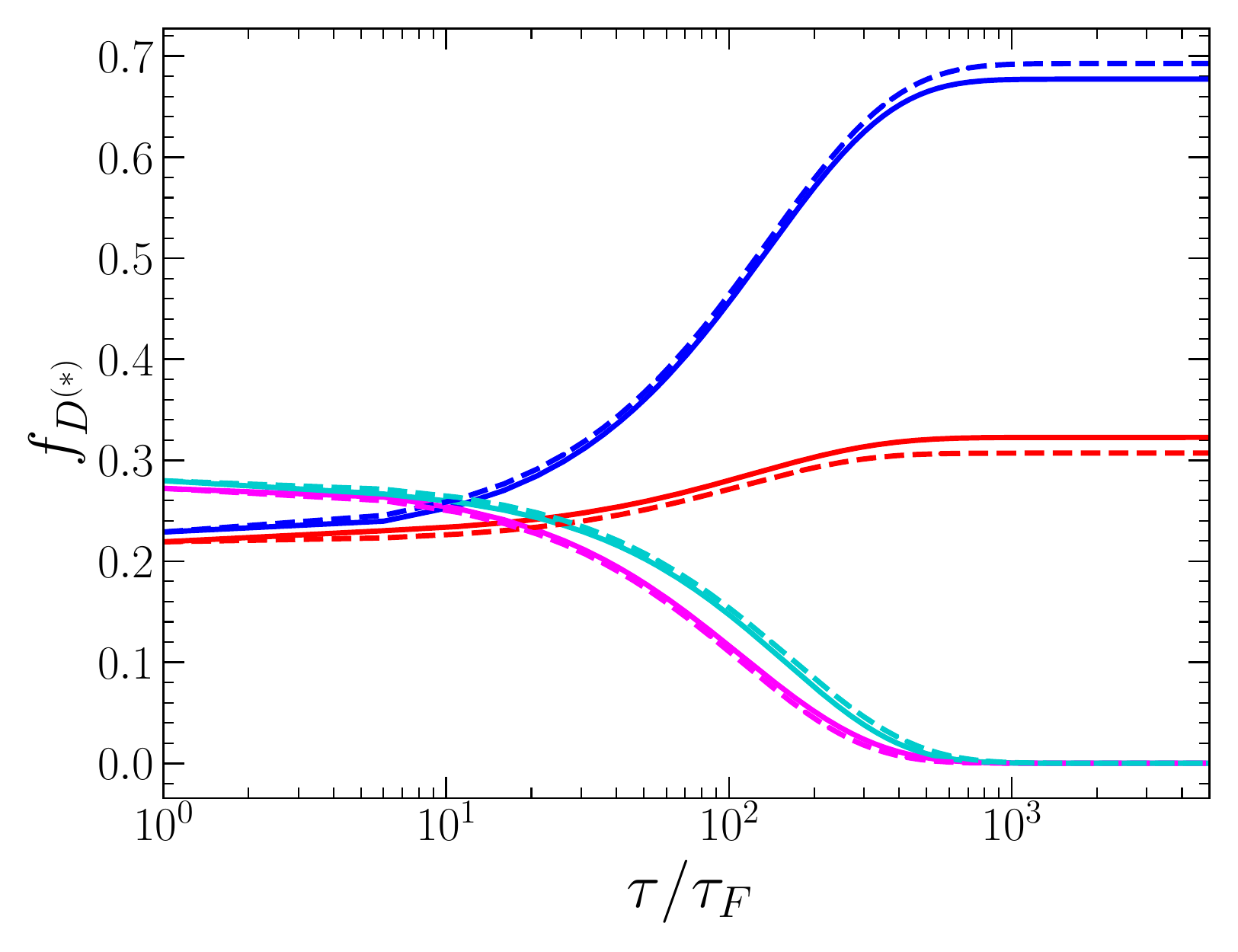}
\caption{
Proper-time evolution of the 
charm-meson fractions $f_{D^0}$ (blue), $f_{D^+}$ (red), $f_{D^{*0}}$ (cyan), and $f_{D^{*+}}$ (magenta)
from solving Eqs.~\eqref{dnD,D*a/dt} (solid curves) 
and from solving Eqs.~\eqref{dnD,D*a/dt} with only the $D^*$ decay terms (dashed curves).
The four fractions add up to 1.
}
\label{fig:density-diff-evol-1}
\end{figure}

To illustrate the evolution of the charm-meson densities, we use Eq.~\eqref{nD,D*-F}, 
with the multiplicities  inferred from SHMc in Eqs.~\eqref{dND0,dNDp},
to provide the initial conditions on $\mathfrak{n}_{D^{(\ast)}}/ \mathfrak{n}_{\pi}$.
If the initial time $\tau$ is identified with the kinetic freezeout time $\tau_F$,
the initial number-density ratios for $D^0$, $D^+$, $D^{\ast 0}$, and $D^{\ast +}$ are 
 0.00276, 0.00264, 0.00337, and 0.00328, respectively.
We define charm-meson fractions by
$f_{D^{(\ast)}} = \mathfrak n_{D^{(\ast)}} /
(\mathfrak{n}_{D^0} + \mathfrak{n}_{D^+} + \mathfrak{n}_{D^{*0}} + \mathfrak{n}_{D^{*+}})$.
The solutions to the evolution equations in Eqs.~\eqref{dnD,D*a/dt} 
are shown in Fig.~\ref{fig:density-diff-evol-1}.
The fractions $f_{D^{*0}}$ and $f_{D^{*+}}$ 
decrease exponentially to 0 on time scales comparable to the $D^\ast$ lifetimes.
The ratio of the multiplicities of $D^0$ and $D^+$ is predicted to
increase from 1.044  at kinetic freezeout to 2.100  at the detector.
If the $t$-channel singularity term in Eq.~\eqref{vsigmapiD*0D*+,+0} is omitted,
the ratio increases instead to 2.168.
The solutions to Eqs.~\eqref{dnD,D*a/dt} 
including only the $D^*$ decay terms with the vacuum values of $\Gamma_{\ast a,b}$ 
are also shown in Fig.~\ref{fig:density-diff-evol-1}.

At times $\tau$ large enough that $n_\pi \ll \Gamma_* f_\pi^4/m_\pi^2$,
 the only terms in Eqs.~\eqref{dnD,D*a/dt} that survive are 
1-body terms with only a single factor of $\mathfrak{n}_D$ or $\mathfrak{n}_{D^\ast}$
and no factors of $\mathfrak{n}_\pi$.
There are 1-body terms from the decays $D^\ast \to D \pi$ and $D^\ast \to D \gamma$.
There are additional 1-body terms from the $t$-channel singularities in $\pi D^* \to \pi D^*$.
The limiting behavior of the reaction rate in Eq.~\eqref{vsigmapiD*0D*+,+0} as $\mathfrak{n}_\pi \to 0$ is
\beq
 \left\langle v\sigma_{\pi {\ast 0} \to \pi {\ast +}} \right\rangle \longrightarrow \frac{1}{3\, \mathfrak{n}_\pi}\,
\frac{\Gamma_{*0,0}\, \Gamma_{*+,0}}{\Gamma_{*0,0} + \Gamma_{*+,0}}.
\eeq
The factor of $1/\mathfrak{n}_\pi$ in the reaction rate cancels the explicit factor of
$\mathfrak{n}_\pi$ in the evolution equation in Eq.~\eqref{dnD*a/dt}.
If we only keep the 1-body terms with the vacuum values of $\Gamma_{\ast a,b}$,
the evolution equations can be solved analytically.
The resulting predictions for the numbers of $D^0$ and $D^+$ are
\begin{widetext}
\begin{subequations}
\begin{align}
N_0 &=
\left(N_0\right)_0
+ \left( 1 - \frac{(1-B_{+0}) \Gamma_{*+}\,  \gamma }{ \Gamma_{*+}\Gamma_{*0}+  (\Gamma_{*+} \!+\!\Gamma_{*0})\, \gamma} \right)  \left( N_{\ast 0} \right)_0
+ \left( B_{+0} + \frac{(1-B_{+0}) \Gamma_{*0}\,  \gamma }{\Gamma_{*+}\Gamma_{*0}+  (\Gamma_{*+} \!+\! \Gamma_{*0})\, \gamma}  \right) 
 \left( N_{\ast +} \right)_0,
\label{ND0-t}
\\ 
N_+ &=
\left(N_+\right)_0
 + \frac{(1-B_{+0})\Gamma_{*+}\,  \gamma }{\Gamma_{*+}\Gamma_{*0}+  (\Gamma_{*+} \!+\! \Gamma_{*0})\, \gamma}  
\left( N_{\ast 0} \right)_0
+ \left( 1-B_{+0} - \frac{(1-B_{+0}) \Gamma_{*0}\,  \gamma }{ \Gamma_{*+}\Gamma_{*0} + (\Gamma_{*+} \!+\! \Gamma_{*0})\, \gamma}  \right) 
 \left( N_{\ast +} \right)_0,
\label{ND+-t}
\end{align}
\label{ND0,+-t}%
\end{subequations}
\end{widetext}
where $1/\gamma = 1/(B_{00}\Gamma_{\ast 0}) + 1/(B_{+0} \Gamma_{\ast +})$,
and $B_{00} = (64.7 \pm 0.9)\%$ is the branching fraction for $D^{\ast 0} \to D^0 \pi^0$.
The coefficients in Eqs.~\eqref{ND0,+-t} depend only on $B_{+0}$, $B_{00}$, and $\Gamma_{\ast 0}/\Gamma_{\ast +}$.
If the multiplicities before $D^*$ decays are those inferred from SHMc  in Eqs.~\eqref{dND0,dNDp},
the ratio $N_0/N_+$ is predicted to increase from 1.044  to  $2.177 \pm 0.016$  at the detector,
where the error bar is from $B_{+0}$, $B_{00}$, $\Gamma_{\ast 0}$, and $\Gamma_{\ast +}$ only.
The difference between this ratio and the naive prediction after Eqs.~\eqref{dND0,dNDp}, 
which ignores $t$-channel singularities, is  $-0.079 \pm 0.006$,
which differs from 0 by about 13 standard deviations. 
 
\textbf{Conclusions.}
We have studied the evolution of charm mesons after the kinetic freeze-out
of an expanding hadron gas produced by a central heavy-ion collision.
We have shown that the $t$-channel singularities in charm-meson reactions 
can have observable consequences.
Their contributions to the reaction rates for $\pi D^\ast \to \pi D^\ast$ 
are inversely proportional to the thermal width of an exchanged $D$,
 which is proportional to $\mathfrak{n}_{\pi}$
because the hadron gas consists predominantly of pions.
The corresponding terms in the evolution equation for charm-meson number densities 
are effectively 1-body terms like those from charm-meson decays.
The conventional predictions for charm-meson ratios in Eqs.~\eqref{ND0,p}
are replaced by those in Eqs.~\eqref{ND0,+-t}.

There are other charm-meson reactions with $t$-channel singularities
including $\pi D^\ast\leftrightarrow \pi\pi D$.
The reaction $\pi D^\ast \to \pi\pi D$ has a pion $t$-channel singularity from the decay
$D^\ast \to D \pi$ followed by the scattering $\pi \pi \to \pi \pi$.
Its reaction rate is higher order in the HH$\chi$EFT expansion than that for $\pi D^\ast \to \pi D^\ast$,
but preliminary results indicate that pion $t$-channel singularities
are more important than $D$-meson $t$-channel singularities.

There have been previous studies of the effects of a thermal hadronic medium on charm mesons
\cite{Fuchs:2004fh,He:2011yi,Montana:2020lfi,Montana:2020vjg}.
In these studies, $t$-channel singularities have been completely  overlooked.
It might be worthwhile to look for other aspects of the thermal physics of charm mesons 
in which the effects of $t$-channel singularities are significant.
One such aspect is the production of the exotic heavy hadrons $X(3872)$ and $T_{cc}^+(3875)$.
Their tiny binding energies relative to a charm-meson-pair threshold imply that they
are loosely bound charm-meson molecules.
In previous studies of the production of charm-meson molecules,   it has been assumed
that they are produced  before kinetic freezeout
\cite{Cho:2013rpa,MartinezTorres:2014son,ExHIC:2017smd,Chen:2021akx,Hu:2021gdg,Abreu:2022lfy,Yoon:2022voo}.
It is possible that there could be significant effects on their production 
from $t$-channel singularities even after kinetic decoupling.

The problem of $t$-channel singularities is an unavoidable aspect of reactions involving unstable particles.
Unstable particles are ubiquitous in hadronic physics. In the Standard Model of particle physics,
the weak bosons and the Higgs are unstable particles.
Most models of physics beyond the Standard Model have unstable particles.
We have identified a simple aspect of charm-meson physics in which the effects of $t$-channel singularities
are significant. 
This provides encouragement  to look for other effects of $t$-channel singularities
in hadronic, nuclear, and particle physics.

\begin{acknowledgments}
 KI would like to thank Ulrich Heinz for many helpful discussions 
during the early stages of this project.
This work was supported in part by the U.S.\ Department of Energy under grant DE-SC0011726, 
by the Ministry of Science, Innovation and Universities of Spain under grant BES-2017-079860,
by the National Natural Science Foundation of China (NSFC) under grant 11905112,
by the Alexander von Humboldt Research Foundation,
and by the  NSFC and the Deutsche Forschungsgemeinschaft (DFG)
through the Sino-German Collaborative Research Center TRR110 
(NSFC grant 12070131001, DFG Project-ID 196253076-TRR110).
\end{acknowledgments}



\begin{thebibliography}{10}

\bibitem{Peierls:1961zz}
R.F.~Peierls,
Possible Mechanism for the Pion-Nucleon Second Resonance,
Phys.\ Rev.\ Lett.\ \textbf{6}, 641-643 (1961).

\bibitem{Melnikov:1996na}
K.~Melnikov and V.G.~Serbo,
New type of beam size effect and the $W$ boson production at $\mu^+ \mu^-$ colliders,
Phys.\ Rev.\ Lett.\ \textbf{76}, 3263 (1996)
[hep-ph/9601221].

\bibitem{Grzadkowski:2021kgi}
B.~Grzadkowski, M.~Iglicki and S.~Mr\'owczy\'nski,
$t$-channel singularities in cosmology and particle physics,
[arXiv:2108.01757].


\bibitem{Burdman:1992gh}
G.~Burdman and J.F.~Donoghue,
Union of chiral and heavy quark symmetries,
Phys.\ Lett.\ B \textbf{280}, 287 (1992).

\bibitem{Wise:1992hn}
M.B.~Wise,
Chiral perturbation theory for hadrons containing a heavy quark,
Phys.\ Rev.\ D \textbf{45}, R2188 (1992).

\bibitem{Cheng:1992xi}
H.Y.~Cheng, C.Y.~Cheung, G.L.~Lin, Y.C.~Lin, T.M.~Yan and H.L.~Yu,
Chiral Lagrangians for radiative decays of heavy hadrons,
Phys.\ Rev.\ D \textbf{47}, 1030 (1993)
[hep-ph/9209262].

\bibitem{Dong:2019byy}
X.~Dong, Y.~J.~Lee and R.~Rapp,
Open Heavy-Flavor Production in Heavy-Ion Collisions,
Ann.\ Rev.\ Nucl.\ Part.\ Sci.\ \textbf{69}, 417-445 (2019)
[arXiv:1903.07709].

\bibitem{Andronic:2005yp}
A.~Andronic, P.~Braun-Munzinger and J.~Stachel,
Hadron production in central nucleus-nucleus collisions at chemical freeze-out,
Nucl.\ Phys.\ A \textbf{772}, 167-199 (2006)
[nucl-th/0511071].

\bibitem{Andronic:2021erx}
A.~Andronic, P.~Braun-Munzinger, M.~K.~K\"ohler, A.~Mazeliauskas, K.~Redlich, J.~Stachel and V.~Vislavicius,
The multiple-charm hierarchy in the statistical hadronization model,
JHEP \textbf{07}, 035 (2021)
[arXiv:2104.12754].

\bibitem{Bjorken:1982qr}
J.D.~Bjorken,
Highly Relativistic Nucleus-Nucleus Collisions: The Central Rapidity Region,
Phys.\ Rev.\ D \textbf{27}, 140-151 (1983)

\bibitem{Hong:2018mpk}
J.~Hong, S.~Cho, T.~Song and S.H.~Lee,
Hadronic effects on the $cc\bar{q}\bar{q}$ tetraquark state in relativistic heavy ion collisions,
Phys.\ Rev.\ C \textbf{98}, 014913 (2018)
[arXiv:1804.05336].

\bibitem{Abreu:2020ony}
L.M.~Abreu,
$X_J(2900)$ states in a hot hadronic medium,
Phys.\ Rev.\ D \textbf{103}, 036013 (2021)
[arXiv:2010.14955].

\bibitem{ALICE:2019hno}
S.~Acharya \textit{et al.} [ALICE],
Production of charged pions, kaons, and (anti-)protons in Pb-Pb and inelastic $pp$ collisions at $\sqrt {s_{NN}}$ = 5.02 TeV,
Phys. Rev. C \textbf{101}, 044907 (2020)
[arXiv:1910.07678].

\bibitem{Gasser:1986vb}
J.~Gasser and H.~Leutwyler,
Light Quarks at Low Temperatures,
Phys.\ Lett.\ B \textbf{184}, 83-88 (1987).


\bibitem{Fuchs:2004fh}
C.~Fuchs, B.~V.~Martemyanov, A.~Faessler and M.~I.~Krivoruchenko,
D-mesons and charmonium states in hot pion matter,
Phys.\ Rev.\ C \textbf{73}, 035204 (2006)
[nucl-th/0410065].

\bibitem{He:2011yi}
M.~He, R.~J.~Fries and R.~Rapp,
Thermal Relaxation of Charm in Hadronic Matter,
Phys.\ Lett.\ B \textbf{701}, 445-450 (2011)
[arXiv:1103.6279].

\bibitem{Montana:2020lfi}
G.~Monta\~na, \`A.~Ramos, L.~Tolos and J.M.~Torres-Rincon,
Impact of a thermal medium on $D$ mesons and their chiral partners,
Phys.\ Lett.\ B \textbf{806}, 135464 (2020)
[arXiv:2001.11877].

\bibitem{Montana:2020vjg}
G.~Monta\~na, \`A.~Ramos, L.~Tolos and J.~M.~Torres-Rincon,
Pseudoscalar and vector open-charm mesons at finite temperature,
Phys.\ Rev.\ D \textbf{102}, no.9, 096020 (2020)
[arXiv:2007.12601].

\bibitem{Cho:2013rpa}
S.~Cho and S.H.~Lee,
Hadronic effects on the $X(3872)$ meson abundance in heavy ion collisions,
Phys.\ Rev.\ C \textbf{88}, 054901 (2013)
[arXiv:1302.6381].

\bibitem{MartinezTorres:2014son}
A.~Martinez Torres, K.P.~Khemchandani, F.S.~Navarra, M.~Nielsen and L.M.~Abreu,
On $X(3872)$ production in high energy heavy ion collisions,
Phys.\ Rev.\ D \textbf{90}, 114023 (2014)
[arXiv:1405.7583].

\bibitem{ExHIC:2017smd}
S.~Cho \textit{et al.} [ExHIC],
Exotic hadrons from heavy ion collisions,
Prog.\ Part.\ Nucl.\ Phys.\ \textbf{95}, 279-322 (2017)
[arXiv:1702.00486].

\bibitem{Chen:2021akx}
B.~Chen, L.~Jiang, X.H.~Liu, Y.~Liu and J.~Zhao,
$X(3872)$ Production in Relativistic Heavy-Ion Collisions,
Phys.\ Rev.\ C \textbf{105}, 054901 (2022)
[arXiv:2107.00969].

\bibitem{Hu:2021gdg}
Y.~Hu, J.~Liao, E.~Wang, Q.~Wang, H.~Xing and H.~Zhang,
Production of doubly charmed exotic hadrons in heavy ion collisions,
Phys.\ Rev.\ D \textbf{104},  L111502 (2021)
[arXiv:2109.07733].

\bibitem{Abreu:2022lfy}
L.~M.~Abreu, F.~S.~Navarra and H.~P.~L.~Vieira,
Multiplicity of the doubly charmed state $T_{cc}^+$ in heavy-ion collisions,
Phys. Rev. D \textbf{105}, 116029 (2022)
[arXiv:2202.10882].

\bibitem{Yoon:2022voo}
H.O.~Yoon, D.~Park, S.~Noh, A.~Park, W.~Park, S.~Cho, J.~Hong, Y.~Kim, S.~Lim and S.H.~Lee,
$X(3872)$ and $T_{cc}$: structures and productions in heavy ion collisions,
[arXiv:2208.06960].


\end{thebibliography}
\end{document}